\documentstyle{article} 
\begin{document}
%---------------------------------------------------------------------------
\title{Mass for the graviton}
%---------------------------------------------------------------------------
\author{
Matt Visser${}^{\dagger}$\\
Physics Department \\
Washington University\\
Saint Louis\\
Missouri 63130-4899\\
USA
}
%---------------------------------------------------------------------------
\date{\small gr-qc/9705051\\
            21 March 1997 \\
            Revised 25 February 1998}
%---------------------------------------------------------------------------
\maketitle
%---------------------------------------------------------------------------
\begin{abstract}
\noindent
Can we give the graviton a mass? Does it even make sense to speak
of a massive graviton? In this essay I shall answer these questions
in the affirmative.  I shall outline an alternative to Einstein
Gravity that satisfies the Equivalence Principle and automatically
passes all classical weak-field tests ($GM/r\approx 10^{-6}$). It
also passes medium-field tests ($GM/r\approx 1/5$), but exhibits
radically different strong-field behaviour ($GM/r\approx 1$).
Black holes in the usual sense do not exist in this theory, and
large-scale cosmology is divorced from the distribution of matter.
To do all this we have to sacrifice something: the theory exhibits
{\em prior geometry}, and depends on a non-dynamical background
metric.
\par\medskip
%----------------------------------------------------------------------------
\noindent
Keywords: graviton, mass.
%----------------------------------------------------------------------------
\par\medskip
\noindent
${}^\dagger$  E-mail: visser@kiwi.wustl.edu \\
web: http://www.wustl.physics.edu/\~{}visser
%----------------------------------------------------------------------------
\par\medskip
\noindent
Based on an essay that was awarded an honorable mention
in the 1997 Gravity Research Foundation essay competition.
%----------------------------------------------------------------------------
\end{abstract}
%----------------------------------------------------------------------------

\newpage
%--------------------------------------------------
\section{Introduction}
%---------------------------------------------------
\def\tr{\hbox{\rm tr}}
\def\implies{\Rightarrow}
%---------------------------------------------------

Can we give the graviton a mass? What would giving the graviton a
mass mean? Does it really make any sense to even speak of a massive
graviton?  These are subtle issues which in the past have led to
considerable confusion. In particular, it is far from clear how to
extrapolate a graviton mass defined for weak fields back into the
strong-field regime.  In this essay I shall show that doing so
entails some surprises.

Recall that there is a general uniqueness result for Einstein
gravity~\cite[pages 417, 429, 431]{MTW}. Any theory of gravity
which:
\begin{enumerate}
\item
is a metric theory, (roughly speaking: satisfies the Equivalence
Principle),
\item
has field equations linear in second derivatives of the metric,
\item
does not have higher-order derivatives in the field equations,
\item
satisfies the Newtonian limit for weak fields,
\item 
and, {\em does not depend on any prior geometry}, (has no background
metric),
\end{enumerate}
must be exactly Einstein gravity itself, thereby implying an exactly
massless graviton. Thus introducing a graviton mass will clearly
require some rather drastic mutilation of the usual foundations
underlying Einstein gravity.  To accommodate a massive graviton
without sacrificing experimental results such as the E\"otv\"os
experiment and the Newtonian limit, and do so without the theoretical
complications of a higher-derivative theory, I shall explore the
option of adding prior geometry by introducing a background metric.

In the weak-field limit ($g_{\mu\nu}= \eta_{\mu\nu} + h_{\mu\nu}$;
$h \ll 1$) the field equations for a massless graviton (in the
Hilbert--Lorentz gauge) are
\begin{equation}
\Delta \left[ h_{\mu\nu} -{1\over2} \eta_{\mu\nu} h \right] +O(h^2) = 
8\pi G \; T_{\mu\nu}.
\end{equation}
We can get this from the action
\begin{equation}
{\cal S} =  \int dx \sqrt{-\eta} \left\{
{1\over2} \left[ h^{\mu\nu} \Delta h_{\mu\nu} - {1\over2} h \Delta h \right] 
+ O(h^3)
- 8\pi G h^{\mu\nu} \; T_{\mu\nu} \right\}.
\end{equation}
In this same limit, it is natural to define the field equations
for a massive graviton to be\footnote{As we shall soon see,
``natural'' is a loaded word in this context.}
\begin{equation}
\label{E-linear}
\Delta \left[ h_{\mu\nu} -{1\over2} \eta_{\mu\nu} h \right]
+ {m_g^2 c^2\over \hbar^2}  
\left[ h_{\mu\nu} -{1\over2} \eta_{\mu\nu} h \right] 
+O(h^2) = 8\pi G \; T_{\mu\nu}.
\end{equation}
The relevant action is now\footnote{Note that the linearized mass
term is {\em not} the Pauli--Fierz term that is the main center of
interest in the Van-Dam--Veltman~\cite{VanDam,VanDam2},
Ford--Van-Dam~\cite{VanDam3}, and Boulware--Deser~\cite{Boulware-Deser}
analyses. This fact is essential to having a well-behaved classical
limit as the graviton mass goes to zero, and will be the topic of
a more extensive forthcoming publication~\cite{Visser}. I wish to
thank Larry Ford for emphasizing the importance of the consistency
problems wrapped up in this issue.}
\begin{eqnarray}
{\cal S} &=&  
\int d^4x \sqrt{-\eta} \Bigg\{
{1\over2} 
\left[ h^{\mu\nu} \Delta h_{\mu\nu} - {1\over2} h \Delta h \right] 
+ {1\over2}{m_g^2 c^2\over \hbar^2}  
\left[ h^{\mu\nu} h_{\mu\nu} - {1\over2} h^2 \right]
\nonumber\\
&&
\qquad\qquad
+ O(h^3)
- 8\pi G h^{\mu\nu} T_{\mu\nu} \Bigg\}.
\end{eqnarray}
The first term is easily extrapolated back to strong fields: it is
simply the quadratic term in the linearization of the usual
Einstein--Hilbert Lagrangian ($\int dx \sqrt{-g} R(g)$). It is the
second term---the mass term for the graviton---that does not have
a clear extrapolation back to strong fields. The key is to introduce
a background metric $g_0$, which will not be subject to a dynamical
equation (at least not classically), and write\footnote{There is
great deal of arbitrariness in writing down the mass term. Any
algebraic function of the metric and background metric that has
the correct linearized behaviour up to second order in $h$ would
do. (See also~\cite{Boulware-Deser}.)}
\begin{eqnarray}
{\cal S}_{mass}(g,g_0) &=& +{1\over2} {m_g^2 c^2\over \hbar^2} \;
\int \sqrt{-g_0} \Bigg\{
(g_0^{-1})^{\mu\nu} \; (g -g_0)_{\mu\sigma} \;
(g_0^{-1})^{\sigma\rho} \;  (g- g_0)_{\rho\nu}
\nonumber\\
&&\qquad\qquad
- {1\over2} [(g_0^{-1})^{\mu\nu} \; (g -g_0)_{\mu\nu}]^2 
\Bigg\}. 
\end{eqnarray}
This mass term depends on two metrics: the dynamical spacetime-metric,
$g$, and the non-dynamical background metric, $g_0$, and makes
perfectly good sense for arbitrarily strong gravitational fields. The
weak-field limit consists of taking $g= g_0 + h$ with $h$ small.

%----------------------------------------------------------------
\section{The model}
%----------------------------------------------------------------

The full action for the variant theory of gravity I will consider
in this essay is\footnote{The original version of this essay
discussed a variant of the current model that was seriously flawed
by internal inconsistency --- that version was actually a variant
of massive Brans--Dicke theory in disguise. I wish to thank David
Garfinkle for pointing out the serious problems in that model.}
\begin{equation}
{\cal S} = \int d^4x 
\left[ 
\sqrt{-g} \;{R(g)\over 16\pi G} + 
\sqrt{-g_0} \; {\cal L}_{mass}(g,g_0) + 
\sqrt{-g} \; {\cal L}_{matter}(g) 
\right].
\end{equation}
Note that the background metric shows up in only one place: in
the mass term for the graviton. The equations of motion for
arbitrarily strong gravitational fields are
\begin{eqnarray}
G^{\mu\nu} &=& 8\pi G \; T^{\mu\nu}
\nonumber\\
&-&
{m_g^2 c^2\over \hbar^2} \;
\left\{
(g_0^{-1})^{\mu\sigma} 
\left[ (g- g_0)_{\sigma\rho} - {1\over2} (g_0)_{\sigma\rho} \;  
(g_0^{-1})^{\alpha\beta} (g- g_0)_{\alpha\beta} \right] 
(g_0^{-1})^{\rho\nu} 
\right\}.
\nonumber\\
\end{eqnarray}
As the mass of the graviton goes to zero we smoothly recover the
ordinary Einstein field equations --- the Lagrangian and field
equations are in this limit both identical to the usual ones. The
only effect, {\em at the level of the field equations}, is to
introduce what is effectively an extra contribution to the
stress-energy\footnote{There is of course also considerable ambiguity
in this effective stress-energy term, and in the strong-field
equations of motion. Any strong-field equation that exhibits the
appropriate linearized behaviour around flat spacetime is a reasonable
candidate for ``massive gravity''. From the point of view espoused
in this essay, anything that linearizes to equation~(\ref{E-linear})
is acceptable.}
\begin{equation}
T_{mass}^{\mu\nu} = 
- {m_g^2 c^2\over 8\pi G \hbar^2}  
\left\{
(g_0^{-1})^{\mu\sigma} \;
\left[ (g- g_0)_{\sigma\rho} - {1\over2} (g_0)_{\sigma\rho} \;  
(g_0^{-1})^{\alpha\beta} (g- g_0)_{\alpha\beta} \right] 
(g_0^{-1})^{\rho\nu}
\right\}.
\end{equation}
The field equations can now be rearranged to look more like the
usual Einstein equations:
\begin{equation}
G^{\mu\nu} =  8\pi G \, \left[ T^{\mu\nu}_{mass} + T^{\mu\nu} \right]. 
\end{equation}

%----------------------------------------------------------------------
\section{Experimental tests: Weak field}
%----------------------------------------------------------------------

To precisely specify the weak-field limit we will have to pick a
particular background geometry for our non-dynamical metric. The
most sensible choice for almost all astrophysical applications is
to take $g_0$ to correspond to a flat spacetime (Minkowski space),
in which case we absorb all of the coordinate invariance in the
theory by going to Cartesian coordinates to make the components of
$g_0$ take on the canonical Minkowski-space values. Once we have
done this there is no further coordinate invariance left. In
particular, it is meaningless to attempt to impose the Hilbert--Lorentz
{\em gauge condition}, which is at first a little puzzling since
we needed the Hilbert--Lorentz condition to set up the linearized
weak field theory in the first place. The resolution to this apparent
paradox is that the conservation of stress-energy implies, among
other things, that
\begin{equation}
\nabla_{\mu} T_{mass}^{\mu\nu} = 0.
\end{equation}
Here $\nabla$ denotes the covariant derivative calculated using
the dynamical metric $g$. If we now linearize this equation around
the non-dynamical metric $g_0$ we find that the Hilbert--Lorentz
condition emerges naturally as a consequence of the equations of
motion, not as a gauge condition. (Exactly the same phenomenon
occurs when we give the photon a small mass via the Proca Lagrangian.
The Lorentz condition, $\partial_{\mu} A^\mu =0$, then emerges as
consequence of electric current conservation, instead of being an
electromagnetic gauge condition.)\footnote{See also the similar
comments in~\cite{Boulware-Deser}.} The analysis of the weak field
limit proceeds in exactly the same way as for ordinary Einstein
gravity.  The gravitational field surrounding a point particle of
mass $M$ and four-velocity $V^\mu$ is approximated at large distances
by\footnote{In obtaining this particular form of the weak-field
metric it is absolutely essential that the mass term I have introduced
is not the Pauli--Fierz term. Again, further details will be deferred
to a forthcoming paper~\cite{Visser}.}
\begin{equation}
\label{E-weak-field}
g_{\mu\nu} \approx 
\eta_{\mu\nu} + {2GM\over r} \exp\left(-{m_g r\over\hbar}\right) 
\left[2 V_\mu V_\nu + \eta_{\mu\nu} \right].
\end{equation}
The only intrinsically new feature here is the exponential Yukawa
fall-off of the field at large distances.

From astrophysical observations the {\bf Particle Data Group} is
currently quoting an experimental limit of~\cite{PDG}
\begin{equation}
m_g < 2 \times 10^{-29} \hbox{ electron--Volts} 
\approx 2 \times 10^{-38} m_{nucleon},
\end{equation}
corresponding to a Compton wavelength of
\begin{equation}
\lambda_g = {\hbar\over m_g c} > 
6\times 10^{22} \hbox{ metres} 
\approx 2 \hbox{ Mega--parsecs}.
\end{equation}
However, insofar as these estimates are based on galactic
dynamics~\cite{Goldhaber,Hare}, the continuing controversies
surrounding the dark-matter/missing-mass problem (relevant already
at distance scales of order kilo-parsecs) should inspire a certain
caution concerning the possibly over-enthusiastic nature of this
limit.  Still, even with an uncertainty of a factor of a thousand
or so in this bound it is clear that the Compton wavelength of the
graviton should be much larger than the dimensions of the solar
system. The relevant exponentials are all well approximated by $1$
for solar system physics, and so this variant theory of gravity
automatically passes all solar system tests of gravity.\footnote{See
also the recent paper by Will for new solar system limits on the
graviton mass~\cite{Will}.}

There will be small (too small to be observable) effects on the
propagation of gravitational waves. The speed of propagation will
be slightly less than that of light, and will depend on frequency,
with
\begin{equation}
v(\omega) 
= 
c \; \sqrt{1-{m_g^2 c^4\over\hbar^2 \omega^2}} 
= 
c \; \sqrt{1-{\lambda^2\over\lambda_g^2}}. 
\end{equation}
For astrophysically relevant frequencies, and given the limit on
the graviton mass, effects due to this phenomenon are too small to
be observable.\footnote{Though Will has recently argued that
there might be measurable effects in the gravity wave chirps due
to black hole coalescence~\cite{Will}.}

%----------------------------------------------------------------------
\section{Experimental tests: Medium field}
%----------------------------------------------------------------------

Although often presented as strong-field tests of Einstein gravity,
the binary pulsar tests~\cite{Taylor,Darmour} are really medium-field
tests ($GM/r\approx 1/5$).  The present theory also automatically
passes all these medium-field tests. This can be seen by working
perturbatively around the Schwarzschild geometry and noting that
the effective contribution to the stress-energy arising from the
graviton mass can be approximated as
\begin{equation}
T_{mass}^{\hat\mu\hat\nu} \approx 
- {\hbar\over \ell_{Planck}^2 \lambda_g^2}  {GM\over r} \times
\left[
\begin{array}{cccc}
4&0&0&0\\
0&0&0&0\\
0&0&0&0\\
0&0&0&0
\end{array}
\right] + O[(GM/r)^2].
\end{equation}
{\em Tricky points:} Here I have written the Schwarzschild geometry
in harmonic coordinates (and started by working in the coordinate
basis). This is needed to be compatible with the Einstein--Lorentz
condition.  In harmonic coordinates the horizon is at $r=GM$. I
then transform to the orthonormal frame attached to the harmonic
coordinates to obtain the physical components of the graviton mass
contribution to the stress-tensor. This is a double perturbation
expansion --- first in the mass of the graviton and secondly in
the field strength. It should only be trusted for $r\ll\lambda_g$
and $r$ greater than and not too close to $M$. To extend this to
the regime $r \approx \lambda_g$ and greater simply make the
substitution $M\to M\exp(-r/\lambda_g)$ to obtain
\begin{equation}
T_{mass}^{\hat\mu\hat\nu} \approx 
- {\hbar\over \ell_{Planck}^2 \lambda_g^2}  
{GM\exp(-r/\lambda_g) \over r} \times
\left[
\begin{array}{cccc}
4&0&0&0\\
0&0&0&0\\
0&0&0&0\\
0&0&0&0
\end{array}
\right] + O[(GM/r)^2].
\end{equation}
This can also be obtained directly from the weak-field solution,
equation (\ref{E-weak-field}).\footnote{Note that this means that
the effective contribution to the stress energy violates the null
energy condition (NEC) and in fact all of the classical energy
conditions. This should not come as a surprise since asymptotically
we demand that the effective gravitational mass of any isolated
system to be an exponentially {\em decreasing} function of distance:
$m(r) \approx M \exp(-r/\lambda_g)$. The only way that this can
happen is by having a negative effective stress energy in the
asymptotic regime. This argument does not necessarily imply that
the NEC violations persist in the strong-field regime.}

Even if we are in medium-strength fields, $GM/r\approx1/5$, the
extreme smallness of the graviton mass, (or equivalently the extreme
largeness of its Compton wavelength), is enough to render this
effective contribution to the stress tensor completely negligible.
In the medium-field regime the spacetime geometry of the spherically
symmetric vacuum solution ($T^{\mu\nu}=0$; $T_{mass}^{\mu\nu}\neq0$)
will not deviate appreciably from the Schwarzschild geometry.  For
the same reason, the production of gravity waves and consequent
orbital decay will not be significantly affected.\footnote{Subtle
non-leading order effects have recently been discussed by
Will~\cite{Will}.}

%----------------------------------------------------------------------
\section{Experimental tests: Strong field}
%----------------------------------------------------------------------

It is in very strong fields $GM/r\approx1$, that the first evidence
of dramatic departure from Einstein gravity arises. I have identified
two areas where the physics is radically altered: black holes and
cosmology. In both these cases the variant of Einstein gravity that
I am describing in this essay is compatible with the current
experimental situation.

%----------------------------------------------------------------------
\subsection{Black holes?}
%----------------------------------------------------------------------
Suppose we look at a spherically symmetric-static spacetime and
write the metric as
\begin{equation}
ds^2 = - g_{tt}(r) dt^2 + g_{rr}(r) dr^2 + 
         R(r)^2 [ d\theta^2 + \sin^2(\theta) d\varphi^2 ].
\end{equation}
[Because we have used up all the coordinate freedom in reducing
$g_0$ to its Minkowski space form we no longer have the freedom to
go to Schwarzschild coordinates by setting $R(r)=r$.] It is now
easy to see that black holes (of the usual type) do not exist in
this theory. Simply take the trace of the equations of motion to
calculate the Ricci scalar
\begin{equation}
{\bf Ricci}  = - 8 \pi G \; T +  
{m_g^2 c^2\over \hbar^2} 
\left\{ 
g_{tt}(g_{tt}-1) + g_{rr}(g_{rr}-1) + 2 {R(r)^2-r^2\over r^2} 
\right\}.
\end{equation}
A normal Schwarzschild-type event horizon, should one exist, is
characterized by the gravitational potential $g_{tt}$ going to zero,
while $g_{rr}$ tends to infinity. But by the field equations, if we
assume finiteness of the stress-energy tensor, this implies that the
Ricci curvature is going to infinity. Thus singularities cannot
be surrounded by event horizons of the usual type---any singularity
that is present in this theory must be either (1) naked and violate
cosmic censorship, or (2) the horizon must be abnormal in the sense
that both $g_{tt}$ and $g_{rr}$ must tend to zero and change sign at
the horizon.

We can deduce roughly where all the interesting physics happens by
working perturbatively around the Schwarzschild geometry and looking
at the effective stress-energy attributable to the presence of a
graviton mass, which becomes Planck scale once
\begin{equation}
r < GM [1+2(\ell_{Planck}/\lambda_g)].
\end{equation}
This will occur in a thin layer, of proper thickness 
\begin{equation}
\delta\ell = \ell_{Planck}\sqrt{GM/\lambda_g},
\end{equation}
located just above where the event horizon would have been if the
graviton mass were exactly zero. Thus there will be a thin layer
near $r=GM$, typically much narrower than a Planck length, where
the geometry is radically distorted away from the Schwarzschild
metric. (Remember that we are in harmonic coordinates.) Even though
the metric is extremely close to Schwarzschild for $r\gg GM$, the
global topology of the maximally-extended spacetime is nowhere near
that of the Kruskal--Szekeres manifold.  Because of the assumed
existence of the flat background metric $g_0$, the maximally-extended
spacetime is in this case topologically $R^4$.

This is compatible with all current observational evidence regarding
the existence of black holes. The current observational data really
only shows the existence of highly compact heavy objects and does
not directly probe the behaviour or existence of the event horizon
itself.

Those aspects of standard black hole physics that do not depend
critically on the precise geometry at or inside the event horizon
will survive in this theory. For instance, most of the Membrane
Paradigm of black hole physics (and the observational consequences
thereof) survives~\cite{Membrane}.  As long as the ``stretched
horizon'' is more than a few $\delta\ell$ above $r=GM$, the
near-field geometry will be indistinguishable from Schwarzschild.

On the other hand, the process of Hawking radiation (semi-classical
black hole evaporation) depends critically on the precise features
of the event horizon. This is one area where we can expect radical
changes from the conventional picture.

The fundamental reason why horizons are so different in this theory
is that with two metrics in the theory, there are now simple scalar
invariants, such as $g_0^{\mu\nu} g_{\mu\nu} = \tr(g_0^{-1} g)$, 
which blow up at the event horizon. Because the non-dynamical
background metric ``knows'' about asymptotic spatial infinity it
carries information down to the horizon to let the theory know {\em in
a local way} that the horizon is a very special place. In Einstein
gravity, absent the non-dynamical metric, there is no local way for
the theory to ``know'' that the horizon is special.

Another interesting side-effect of the existence of prior geometry
is that the object which in standard Einstein gravity is called
the stress-energy {\em pseudotensor} of the gravitational field
can now be elevated to the status of a true tensor object. This
permits us to now assign a well-defined notion of stress-energy to
the gravitational field itself.

%------------------------------------------------------------
\subsection{Cosmology?}
%------------------------------------------------------------

A second situation in which a small mass for the graviton can have
big effects is in cosmology: The fundamental physics is that with
the Yukawa fall-off providing a long distance cutoff on the
inverse-square law the motion of galaxies separated by more than
a few Compton wavelengths becomes uncorrelated and the large-scale
expansion of the universe is no longer dependent on the cosmological
distribution of matter.

In a cosmological setting it is no longer obvious that we should
use the flat-space Minkowski metric as background. I will keep the
discussion general by using the usual assumed symmetry properties
to deduce that the dynamical metric and non-dynamical metric should
both be Friedmann--Robertson--Walker. If we put the physical metric
into the canonical proper-time gauge
\begin{equation}
ds^2 = -dt^2 + a^2(t) \;  g_{ij} \, dx^i dx^j,
\end{equation}
then we no longer have full freedom to do so with the non-dynamical
background metric and must be satisfied by taking
\begin{equation}
ds^2_0 = - b_0^2(t) dt^2 +  a_0^2(t) \; g_{ij} \, dx^i dx^j.
\end{equation}
Here $b_0(t)$ and $a_0(t)$ are (for the time being) arbitrary
functions of cosmological time $t$. The graviton mass term in the
effective stress energy tensor is
\begin{equation}
T_{mass}^{\hat\mu\hat\nu} \approx 
{\hbar\over \ell_{Planck}^2 \lambda_g^2}  
\left\{ \eta^{\hat\mu\hat\nu} + {1\over2 a_0^2 b_0^2}
\left[
\begin{array}{cccc}
3 a^2 b_0^2 - a_0^2&0&0&0\\
0&- a^2 b_0^2 - a_0^2&0&0\\
0&0&- a^2 b_0^2 - a_0^2&0\\
0&0&0&- a^2 b_0^2 - a_0^2
\end{array}
\right] \right\}.
\end{equation} 
If we treat $a_0$ as a completely arbitrary function of $t$ then
$b_0$ is determined (as a function of $a_0$ and $a$) via stress-energy
conservation. However this leaves us with a completely arbitrary
contribution to the cosmological stress-energy. That is: an {\em
arbitrary} background geometry, $g_0$, can be used to drive an {\em
arbitrary} expansion for the physical metric, $g$. Consequently
the expansion of the universe is completely divorced from the
cosmological distribution of matter {\em unless we place some
constraints on the choice of background geometry}.

One particularly attractive choice of cosmological background is
the Milne universe~\cite[pages 198--199]{Peebles}.  This consists
of a spatially open universe with $b_0$ constant and $a_0(t) = b_0
\, c t$. Remarkably, this is just flat Minkowski space in disguise,
and in this sense even cosmology can be performed with a flat
background.

A second attractive choice of cosmological background is the
de~Sitter universe~\cite[pages 77--78, 307--310]{Peebles}.  This
consists of a spatially flat universe with $b_0$ constant and
$a_0(t) = b_0 \, \exp(\kappa t)$.

There are many options for the theoreticians to explore, in that
the combination of choosing a background geometry and graviton mass
can potentially influence many standard cosmological tests (primordial
nuclear abundances, cosmic microwave fluctuations, etc.)

Observational cosmologists might like to view this as an opportunity
to feel justified in measuring $a(t)$ directly from the observational
data without interference from theoretical prejudices of how $a(t)$
should behave in normal Einstein gravity. Once $a(t)$ has been
measured, it can be inserted into the Einstein equations to determine
$T_{mass}^{\mu\nu}$. With some independent estimate for $m_g$ we
could then deduce the geometry of the background spacetime $g_0$
by inference from the observational data.

In particular, this is one way of fixing the age-of-the-oldest-stars
problem currently afflicting observational cosmology. (This is by
no means the most attractive solution, attributing the current
crisis to observational error or to a non-zero cosmological constant
are less radical and more attractive solutions.)

%-------------------------------------------------------------------
\section{Discussion}
%-------------------------------------------------------------------

The variant theory of gravity I have sketched in this essay---a
specific proposal for giving the graviton a mass---passes all
present tests of classical gravity.  In fact, since we have more
free variables to play with, it is in better agreement with empirical
reality than the current theory. This should be balanced against
the fact that with enough free parameters we can fit almost anything.
The most interesting part of the theory is that it radically changes
ideas concerning black holes and cosmology---but does so in a way
that is compatible with what we currently know.

The most disturbing part of the theory is the role of the non-dynamical
background metric. For asymptotically flat spacetimes it seems
clear that the appropriate background metric to take is flat
Minkowski space. For cosmological situations the issue is less
clear-cut but the choice of the Milne universe or de~Sitter universe
for the background geometry seems particularly appealing.

Clearly, the theory presented in this essay is far from being
completely and definitively understood: there are a lot of issues
(such as quantization~\cite{VanDam,VanDam2,VanDam3,Boulware-Deser,%
Goldhaber,Hare}) ripe for further development.~\footnote{From the
point of view of Van Dam and Veltman~\cite{VanDam,VanDam2}, Ford
and Van Dam~\cite{VanDam3}, and Boulware and Deser~\cite{Boulware-Deser}
the mass term I have discussed in this note is viewed as pathological
due to an unboundedness of the energy. However this unboundedness
is not something particular to the mass term itself but is merely
a manifestation of the well-known instability that formally afflicts
even the kinetic terms. Thus I would argue that the instability in
the mass term is no worse than the known (formal) instability in
the kinetic terms, and amenable to similar treatment.  I plan to
develop these issues more fully in~\cite{Visser}.} What is particularly
intriguing here is the fact that asking such a simple and basic
question can lead to such unexpected surprises---classical gravity
still exhibits a great potential for confounding the unwary.

Finally, I would be remiss in not mentioning related work of the
Russian school, such as that of Logunov and
co-workers~\cite{Logunov86,Logunov88,Logunov89,Logunov92a,%
Logunov92b,Logunov97}, and that of Loskutov~\cite{Loskutov91a,Loskutov91b}.
Additionally, there have also been attempts at deriving and
calculating a graviton mass from first principles using fundamental
string theory~\cite{Kostalecky-Samuel}.

%------------------------------------------------------------------
\section*{Acknowledgements}
%------------------------------------------------------------------

This work was supported by the U.S. Department of Energy. I wish
to thank David Garfinkle for bringing to my attention serious flaws
in the previous version of this essay, Larry Ford for emphasizing
the subtlety of the zero mass limit, and Cliff Will for his comments
on the experimental situation.

%----------------------------------------------------------------------

%-----------------------------------------------------------------------
\end{document}